\documentclass[aps,pre,twocolumn]{revtex4}

\usepackage{amssymb,amsmath,graphicx}

\begin{document}

\title{Bayesian approach to clustering real value, hypergraph and 
bipartite graph data:\\ solution via variational methods}

\author{Alexei Vazquez\\
The Simons Center for Systems Biology\\
Institute for Advanced Study, Einstein Drive, Princeton, New Jersey 08540, 
USA}

\date{\today}

\begin{abstract}

Data clustering, including problems such as finding network communities, 
can be put into a systematic framework by means of a Bayesian approach. 
The application of Bayesian approaches to real problems can be, however, 
quite challenging. In most cases the solution is explored via Monte Carlo 
sampling or variational methods. Here we work further on the application 
of variational methods to clustering problems. We introduce generative 
models based on a hidden group structure and prior distributions. We 
extend previous attends by Jaynes, and derive the prior distributions 
based on symmetry arguments. As a case study we address the problems of 
two-sides clustering real value data and clustering data represented by a 
hypergraph or bipartite graph. From the variational calculations, and 
depending on the starting statistical model for the data, we derive a 
variational Bayes algorithm, a generalized version of the expectation 
maximization algorithm with a built in penalization for model complexity 
or bias. We demonstrate the good performance of the variational Bayes 
algorithm using test examples.

\end{abstract}

\maketitle

\bibliographystyle{apsrev}

\section{Introduction}

Mixture models provide an intuitive statistical representation of datasets 
structured in groups, clusters or classes \cite{maclachlan00}. A complex 
dataset is decomposed into the superposition of simpler datasets. The 
inverse problem consists in determining the group decomposition and the 
statistical parameters characterizing each group. For a fixed number of 
groups the expectation maximization (EM) algorithm provides a recursive 
solution to the inverse problem \cite{dempster77}. The estimation of the 
right number or groups has been, however, a great challenge. Corrections 
such as the Arkaike information criterion (AIC) \cite{akaike74} and the 
Bayesian information criterion (BIC) \cite{schwarz78} have been derived, 
penalizing model complexity and overfitting. Yet, the number of groups 
estimated from these criteria is in general unsatisfactory.

In contrast, a Bayesian approach would not attempt to estimate what is the 
``optimal'' number of groups, but instead average over models with a 
different number of groups \cite{jeffreys39}. The Bayesian approach is 
becoming a popular technique to solve problems in data analysis, model 
selection and hypothesis testing \cite{spirtes00,mackay03,robert07}. Many 
of the original ideas come from the early work of Jeffreys 
\cite{jeffreys39}, but it is just recently that they are starting to be 
used widely \cite{spirtes00,mackay03,robert07}. The application of 
Bayesian approaches to real problems can be, however, quite challenging. 
In most cases the solution is explored via Monte Carlo sampling 
\cite{chen00,mackay03} or variational methods 
\cite{mackay03,beal03,yedidia05}. The application of variational methods 
to Bayesian problems results in the variational Bayes (VB) algorithm 
\cite{mackay03,beal03}. The VB algorithm is a set of self-consistent 
equations analog to the EM algorithm. They can be solved recursively 
obtaining an approximate solution to the inverse inference problem. These 
methods have been applied, for example, to Gaussian mixture models for 
real value data \cite{maclachlan00,rasmussen00}, Dirichlet mixture models 
for categorical data \cite{blei03} and the problem of finding graph 
modules \cite{hofman07}.

Here we further study the use of variational methods in the context of 
Bayesian approaches, focusing on data clustering problems. I the first two 
sections we review the Bayesian approach. In Section \ref{variational} we 
revisit the connection between the Bayesian formulation and statistical 
mechanics. In section \ref{models} we introduce the generalities of 
generative models with a hidden structure at the samples side and at both 
the samples and variables side. In Section \ref{S:priors} we extend the 
previous work by Jaynes \cite{jaynes68} deriving prior distributions based 
on symmetry properties. We report a correction to his result for the model 
with a location and scale parameter and an extension of his result for the 
binomial model to the multinomial model. In the following Sections we 
study the problem of two-sides clustering real value data and of 
clustering data represented by a hypergraph or bipartite graph. Depending 
on our starting statistical model, we obtain a VB algorithm. Because of 
its Bayesian root, the VB algorithms have a built in correction for model 
complexity or bias and, therefore, they do not require the use of 
additional complexity criteria. The performance of the VB algorithms is 
tested in some examples, obtaining satisfactory results whenever there is 
a significant distinction between the groups.
 
\section{Bayesian approach and variational solution}\label{variational}

The {\em Bayesian approach} is a systematic methodology to interpret 
complex datasets and to evaluate model hypothesis. Its main ingredients or 
steps are: given a dataset $D$, (i) introduce a statistical model with 
model parameters, $\phi$, (ii) write down the likelihood to observe the 
data given the proposed model and parameters, $P(D|\phi)$, (iii) determine 
the prior distribution for the model parameters based on our current 
knowledge, $P(\phi)$, and, finally, (iv) invert the statistical model of 
the data given the likelihood and prior distribution to obtain the 
posterior distribution of the model parameters given the model and data, 
$P(\phi|D)$. The latter step is based on Bayes rule

\begin{equation}\label{Bayes-Theorem}
P(\phi|D) = \frac{1}{Z} P(D|\phi)P(\phi)
\end{equation}

\noindent where 

\begin{equation}\label{Z}
Z = P(D)=\int d\theta P(D|\phi)P(\phi)\ .
\end{equation}

\noindent Having obtained the distribution of the model parameters, at 
least formally, we can determine other magnitudes. For example, the 
average of a quantity $A(\phi)$ is given by

\begin{equation}\label{Ave}
\langle A(\phi)\rangle = \int d\phi P(\phi|D) A(\phi)\ .
\end{equation}

\noindent In practice calculating (\ref{Z}) or (\ref{Ave}) is a formidable 
task. A very powerful approximation scheme is the {\em variational method} 
\cite{mackay03,beal03}. The main idea of the variational method is to 
approximate the generally difficult to handle distribution $P(\phi|D)$ by 
a distribution $Q(\phi|D)$ of a more tractable form. In the following we 
omit the dependency of $Q$ on $D$ and just write $Q(\phi)$. Given 
$Q(\phi)$ we can obtain a bound for $F=-\ln Z$ using Jensen's inequality

\begin{eqnarray}\label{Jensen}
F & = & -\ln Z
\nonumber\\
 & = & -\ln \int d\phi Q(\phi) 
\frac{P(D|\phi)P(\phi)}{Q(\phi)}
\nonumber\\
 & \leq & -\int d\phi Q(\phi) 
\ln \frac{P(D|\phi)P(\phi)}{Q(\phi)}
\end{eqnarray}

\noindent The latter equation can be rewritten as \cite{mackay03}

\begin{equation}\label{F}
F \leq U - TS
\end{equation}

\noindent where $T=1$,

\begin{equation}\label{U}
U = -\int d\phi Q(\phi) \ln P(D|\phi)
\end{equation}

\noindent is minus the average log likelihood and

\begin{equation}\label{S}
S = - \int d\phi Q(\phi) \ln 
\frac{Q(\phi)}{P(\phi)}
\end{equation}

\noindent is the Kullback-Leibler divergence of $Q(\phi)$ relative to the 
prior distribution $P(\phi)$ \cite{kullback59}. Equation (\ref{F}) 
resembles the usual free energy in statistical mechanics: $F = U - TS$, 
where $U$, $S$ and $T$ are the internal energy, entropy and temperature of 
the system, the temperature being expressed in units of the Boltzman 
constant $k_{\rm B}$. Minus the average log likelihood plays the role of 
the internal energy, the Kullback-Leibler divergence of $Q(\phi)$ plays 
the role of the entropy and temperature equals one.

Equation (\ref{F}) emphasizes the two components determining the best 
choice of variational distribution $Q(\phi)$:  better fit to the data and 
model bias. How well the data is fitted is quantified by the internal 
energy $U$ (\ref{U}). To achieve the best fit, or internal energy ground 
state, $Q(\phi)$ should be concentrated around the regions of the 
parameter space where $P(D|\phi)$ is maximum. The best choice in this 
respect will by the maximum likelihood estimate (MLE)

\begin{equation}\label{QMLE}
Q_{\rm MLE}(\phi) = \delta(\phi-\phi^*)
\end{equation}

\noindent where

\begin{equation}\label{phiMLE}
\phi^*=\max_{\phi}P(D|\phi)\ .
\end{equation}

\noindent In the opposite extreme, when no data is presented to us, the 
best distribution is that maximizing the Kullback-Leibler divergence 
relative to the prior distribution. This maximum entropy (ME) solution is 
the prior distribution itself

\begin{equation}\label{QME}
Q_{ME}(\phi) = P(\phi)\ .
\end{equation}

\noindent In general, the drive to better fit the data is opposed by the 
tendency to obtain the least unbiased model. The variational solution is 
therefore in the middle between the one extreme of biased models fitting 
the data very well and completely unbiased models giving a bad fit to the 
data. It is obtained after minimizing (\ref{F}) with respect to $Q(\phi)$ 
over a restricted class of functions. This variational solution $Q(\phi)$ 
represents the closest distribution to $P(\phi|D)$ within the class of 
functions considered.

\section{Statistical model with a population structure}\label{models}

In this section we present the generalities of statistical models with a 
first level population structure. Similar models has been studied in 
\cite{blei03,hofman07}. Our working hypothesis is that there is 
a hidden population structure, characterized by the subdivision of the 
population samples into groups. We assume that we are given a dataset $D$ 
which, in some way to be determined, reflects the population structure. 
The problem consist in inferring this hidden structure and the associated 
model parameters from the data. To tackle this problem we introduce a 
statistical model with a built in population structure as a generative 
model of the data. The population structure and the model parameters are 
then inferred solving the inverse problem. More precisely

\begin{itemize}

\item[(i)] We consider a population composed of $n$ elements divided in 
$K$ groups.

\item[(ii)] The samples assignment to groups is generated by a multinomial 
model with probabilities $\pi_k$, $k=1,\ldots,K$. Denoting by $g_i$ the 
group to which the $i$-th sample belongs, we obtain

\begin{equation}\label{Pgpi}
P(g|\pi) = \prod_{i=1}^n\pi_{g_i}\ .
\end{equation}

\item[(iii)] Given the group assignments $g_i$, and depending on the 
dataset, we write down the likelihood $P(D|g,\theta)$ to observe the data 
parametrized by the parameter set $\theta$.

\item[(iv)] Putting all this together we obtain the posterior distribution

\begin{equation}\label{GM1}
P(\phi|D) = \frac{1}{Z} P(D|g,\theta)P(g|\pi)P(\theta)P(\pi)P(K)\ ,
\end{equation}

\noindent where $\phi=(g,\theta,\pi,K)$ and $P(\theta)$, $P(\pi)$ and 
$P(K)$ are the prior distributions of $\theta$, $\pi$ and $K$.

\end{itemize}

\noindent The form of the prior distributions, except for $P(K)$, is the 
subject of the next section. The distribution $P(K)$ is irrelevant for 
problems with large datasets. The difference between the log-likelihood of 
models with different values of $K$ is in general of the order of the 
dataset size and, as a consequence, the contribution of $\ln P(K)$ is 
negligible. Thus, in the following sections we simply neglect the 
contribution given by $P(K)$. Finally, we specify the likelihood 
$P(D|g,\theta)$ when addressing specific problems.

In some cases we are going to assume that the variables in our dataset are 
also divided in groups. Here we consider a set of $m$ variables divided in 
$L$ groups. The variables assignment to groups is generated by a 
multinomial model with probabilities $\kappa_l$, $l=1,\ldots,L$. Denoting 
by $c_j$, $j=1,\ldots,m$, the variable group to which variable $j$ belongs 
we can then write

\begin{equation}\label{Pckappa}
P(c|\kappa) = \prod_{j=1}^m \kappa_{c_j}
\end{equation}

\noindent After adding this variable group structure, the posterior 
distribution (\ref{GM1}) is replaced by

\begin{eqnarray}\label{GM2}
P(\phi|D) &=& \frac{1}{Z} P(D|g,c,\theta)P(g|\pi)P(c|\kappa)
\nonumber\\
&\times& P(\pi)P(\kappa)P(\theta)P(K)\ ,
\end{eqnarray}

\noindent where $\phi=(g,c,\theta,\pi,\kappa,K)$ and $P(\kappa)$ is the 
prior distribution of $\kappa$.

\section{Prior distributions}\label{S:priors}

\begin{table*}\label{priors}
\begin{tabular}{|l|l|l|l|l|}
%\display
\hline
Model & Likelihood & Conjugate prior & Invariant prior & Renormalization limit\\
\hline
Binomial & $\binom{N}{n} p^n(1-p)^{N-n}$ 
& $ {\rm Beta}(p;\tilde{\alpha},\tilde{\beta}) =
\frac{ 1 }{ {\rm B}(\tilde{\alpha},\tilde{\beta}) }
p^{\tilde{\alpha}-1}(1-p)^{\tilde{\beta}-1} $
& ${\rm const.} p^{-1}(1-p)^{-1}$
& $\tilde{\alpha}\rightarrow0$, $\tilde{\beta}\rightarrow0$\\
Multinomial & $ \frac{ (\sum_{k=1}^Kn_k)! }{ \prod_{k=1}^K n_k! }
\prod_{k=1}^K\pi_k^{n_k}$ 
& $ {\rm D}(\pi;\gamma) =
\frac{1}{{\rm B}(\tilde{\gamma})} \prod_{k=1}^K\pi_k^{\tilde{\gamma_k}-1} $ 
& $ {\rm const.} \prod_{i=1}^K\pi_i^{-1}$
& $\tilde{\gamma}_k\rightarrow0$\\
Normal & $\prod_{i=1}^n 
\frac{ 1 }{ \sqrt{2\pi\sigma^2} } e^{ - \frac{ (X_i-\mu)^2 }{ 2\sigma^2 } }$
& $\frac{ 2\left(\frac{\tilde{\alpha}}{2}\tilde{\sigma}^2\right)^{ \frac{\tilde{\alpha}}{2} } }
{ \Gamma\left(\frac{\tilde{\alpha}}{2}\right)\sigma^{\tilde{\alpha}+1} }
e^{ -\frac{ \tilde{\alpha}\tilde{\sigma}^2 }{ 2\sigma^2 } }
\sqrt{ \frac{ \tilde{\alpha} }{ 2\pi\sigma^2 } }
e^{ -\frac{ \tilde{\alpha}(\mu-\mu_0)^2 }{ 2\sigma^2 } }$
& $\frac{\rm const.}{\sigma^2}$
& $\tilde{\alpha}\rightarrow0$\\
\hline
\end{tabular}

\caption{{\bf Prior distributions:} Examples of model likelihoods and 
their associated conjugated priors and invariant priors. ${\rm 
Beta}(x;a,b)$ denotes the probability density function of the beta 
distribution, where ${\rm B}(a,b)$ is the beta function. ${\rm 
D}(x;\gamma)$ denotes the probability density function of the Dirichlet 
distribution, the generalized beta distribution, where ${\rm 
B}(\gamma)=\Gamma(\sum_k\gamma_k)/\prod_k\Gamma(\gamma_k)$ is the 
generalized beta function. The renormalization limit column indicates the 
limit in which the conjugate prior approaches the invariant prior.}

\end{table*}

The choice of the prior distribution $P(\phi)$ is probably one of the less 
obvious topics in Bayesian analysis. Currently the predominant choice is 
the use of conjugate priors. The form of conjugate priors is indicated by 
the likelihood, making the prior selection less ambiguous. For example, the 
binomial likelihood $P(n|p)\propto p^n (1-p)^{N-n}$ suggests a beta 
distribution for $P(p|n)$. Furthermore, by choosing a beta distribution as 
a prior, $P(p) \propto p^{\tilde{\alpha}-1}(1-p)^{\tilde{\beta}-1}$, the 
posterior distribution remains a beta distribution, but with exponents 
$\alpha=\tilde{\alpha}+n\nonumber$ and $\beta=\tilde{\beta}+N-n$. In this 
sense, the beta distribution is the conjugate prior of the binomial 
likelihood. A list of conjugate priors relevant for this work is provided 
in Table \ref{priors}.

Yet, the fact that the form of conjugate priors is suggested by the 
likelihood does not demonstrate that they are the correct choice of 
priors. Moreover, even if we accept their use, it is not clear what is the 
correct choice for the prior distribution parameters, e.g. 
$\tilde{\alpha}$ and $\tilde{\beta}$. Different methods have been 
proposed to determine these parameters. In general they are based on {\it 
a posteriori} analyzes, e.g. calculations, making use of the data in some 
way or another. Such methods violate, however, the concept of prior 
distribution, defined as the distribution of the model parameters in the 
absence of the data.

An alternative approach is that by Jaynes \cite{jaynes68}. According to 
Jaynes, in the absence of any data, the priors should be solely determined 
based on the symmetries and constraints of the problem under 
consideration. In this work we make use of Jaynes's approach to determine 
the prior distribution. Below we derive Jaynes's priors for the cases 
relevant for this work.

\subsection{Prior for a model with location and scale parameters}\label{LP}

Consider a problem where the data consists of equally distributed random 
variables $X_i$, $i,\ldots,n$, taking real values. Furthermore let us 
assume that the likelihood has the form

\begin{equation}\label{l1}
P(X|\mu,\sigma) = \prod_i f\left( \frac{X_i-\mu}{\sigma} \right) \frac{1}{\sigma}\ ,
\end{equation}

\noindent where $f(x)$ is a probability density function in the real line 
and $\mu$ and $\sigma$ are a location and scale parameter respectively. 
Our task consist in determining the prior distribution of $\mu$ and 
$\sigma$. Now, suppose $X_i$ represent positions, which could be measured 
from difference systems of reference and using different units. In this 
context the prior distribution should be the same regardless of our system 
of reference and units. More precisely, our system is invariant under the 
transformations

\begin{eqnarray}\label{t1}
x^\prime &=& a(x+b)\nonumber\\
\mu^\prime &=& a(\mu+b)\nonumber\\
\sigma^\prime &=& a\sigma
\end{eqnarray}

\noindent where $b$ represents a translation and $a$ a change of scale or 
units. The likelihood is invariant under these transformations and so must 
be the prior distribution. Therefore,

\begin{equation}\label{i1}
P(\mu^\prime,\sigma^\prime) d\mu^\prime d\sigma^\prime = P(\mu,\sigma) d\mu d\sigma
\end{equation}

\noindent The solution to this functional equation is

\begin{equation}\label{p1}
P(\mu,\sigma) = \frac{\rm const.}{\sigma^2}\ .
\end{equation}

\noindent This analysis was first reported by Jaynes \cite{jaynes68}. He 
obtained, however, $P(\mu,\sigma)\propto 1/\sigma$. This discrepancy is 
rooted in the fact that Jaynes did not take into account that the location 
parameter $\mu$ follows the same rules than $x$ upon the translation and 
scale transformations. He assumed $\mu^\prime=\mu+b$ \cite{jaynes68} while 
the correct transformation is $\mu^\prime=a(\mu+b)$ (\ref{t1}).

\subsection{Prior for the multinomial model}\label{PM}

Consider the multinomial model with $K$ states

\begin{equation}\label{l2}
P(n|\pi) =\frac{ \left( \sum_{k=1}^Kn_k \right)! }{ \prod_{i=1}^K\pi_k }
\prod_{k=1}^K \pi_k^{n_k}\ ,
\end{equation}

\noindent where $n_k$ is the number of times state $k$ was observed and 
$\pi_k$ is the probability to observe state $k$ in one trial, 
$0\leq\pi_k\leq1$ and $\sum_{k=1}^K\pi_k=1$. Here we extend the approach 
followed by Jaynes for the binomial model \cite{jaynes68}.

The probabilities $\pi_k$ may be different depending on our believe, e.g. 
all states are equally probable. Different investigators may have 
different believes, resulting in different choices of $\pi_k$. The main 
assumption is that the prior distribution should be independent of what is 
our specific believe and, therefore, should be invariant under a believe 
transformation.

{\em Believe transformation:} Let us represent by $S_k$ the state $k$, and 
let $P(S_k|E)$ and $P(S_k|E^\prime)$ be the probabilities to observe state 
$S_k$ in one trial according to believe $E$ and $E^\prime$, respectively. 
From Bayes rule it follows that

\begin{equation}\label{t2}
P(S_k|E^\prime) = \frac{ P(E^\prime|S_k,E)P(S_k|E) }{ \sum_j 
P(E^\prime|S_j,E)P(S_j|E) }
\end{equation}

\noindent for $k=1,\ldots,K$. The latter equation can be rewritten as

\begin{equation}\label{t3}
\pi_k^\prime = \frac{a_k}{A} \pi_k
\end{equation}

\noindent for $k=1,\ldots,K-1$ and 
$\pi_K^\prime=1-\sum_{k<K}\pi_k^\prime$, where $\pi_k=P(S_k|E)$, 
$\pi_k^\prime=P(S_k|E^\prime)$,

\begin{equation}\label{ai}
a_k = \frac{ P(E^\prime|S_k,E) }{ P(E^\prime|S_K,E) }
\end{equation}

and

\begin{equation}\label{A}
A = 1 + \sum_{k<K} (a_jk-1)\pi_k\ .
\end{equation}

\noindent Equation (\ref{t3}) provides the transformation rules of the 
probabilities $\pi_k$ from one system of believe to another.

The invariance under the above transformation lead to the functional equation 

\begin{equation}\label{i2}
P(\pi^\prime) d\pi^\prime = P(\pi) d\pi\ ,
\end{equation}

\noindent To solve this equation we first need to compute the determinant 
of the transformation Jacobian. The Jacobian of the transformation 
(\ref{t3}) has the matrix elements

\begin{equation}\label{Jij}
J_{ij} = \frac{\partial\pi_i^\prime}{\partial\pi_j} 
= \frac{a_i\delta_{ij}}{A} - \frac{a_i(a_j-1)\pi_i}{A^2}\ ,
\end{equation}

\noindent $i,j=1,\ldots,K-1$. This matrix can be decomposed into the 
product $J=BC$, where $B_{ij}=a_i\delta_{ij}/A$ is a diagonal matrix and 
$C_{ij}=\delta_{ij}-(a_j-1)\pi_i/A$ has two eigenvalues, 
$\lambda_1=A^{-1}$ and a $n-2$-degenerate eigenvalue $\lambda_2=1$. 
Putting all together we obtain

\begin{equation}\label{dJ}
|J| = |B|\lambda_1\lambda_2^{n-2} = \frac{1}{A^n} \prod_{k=1}^K a_k\ .
\end{equation}

\noindent The solution of (\ref{i2}), with $d\pi^\prime = |J|d\pi$, is 
given by

\begin{equation}\label{p2}
P(\pi) = {\rm const.} \prod_{i=1}^K \pi_i^{-1}\ .
\end{equation}

\noindent Note that for $K=2$, $\pi_1=p$ and $\pi_2=1-p$, we recover the 
result by Jaynes for the binomial model

\begin{equation}\label{p3}
P(p)\propto p^{-1}(1-p)^{-1}\ .
\end{equation}

\subsection{Improper priors renormalization}

The prior distributions (\ref{p1}) and (\ref{p2}) are improper, i.e. their 
integral over the parameter space is not finite. At first this may sound 
an unsuitable property for a prior distribution. Nevertheless, the 
improper nature of these prior distributions is just indicating that the 
symmetries in our problem are not sufficient to fully determine them. Data 
is required to obtain a proper distribution. The best example for an 
intuitive understanding of these arguments is the prior distribution of 
the location parameter. In the absence of any data and under the 
assumption of translational invariance, it is clear that every value in 
the real line is an equally probable value for the location parameter, 
resulting in an improper prior.

From the operational point of view, the posterior distribution may be 
proper even when the prior is not. Indeed, the 
integral $\int d\phi 
P(\phi)$ may be improper, $\int d\phi P(\phi|D) \propto \int d\phi 
P(D|\phi)P((\phi)$ may be proper. The posterior distribution can be 
improper when the inference problem has not been correctly formulated or 
there is not sufficient data to determine the model parameters.

To avoid dealing with improper distributions, we can renormalize improper 
priors to some limit of a proper distribution. Since conjugate priors 
facilitate analytical calculations they are a good starting point. This is 
illustrated in Table (\ref{priors}) for selected examples. These are the 
prior distributions used herein. In particular, for the multinomial 
probabilities $\pi$ and $\kappa$ we use the renormalized invariant priors

\begin{equation}\label{Ppi}
P(\pi) = \frac{1}{{\rm B}(\tilde{\gamma})} 
\prod_{k=1}^K \pi^{\tilde{\gamma}_k-1}
\end{equation}

\begin{equation}\label{Pkappa}
P(\kappa) = \frac{1}{{\rm B}(\tilde{\epsilon})} 
\prod_{l=1}^L \kappa^{\tilde{\epsilon}_l-1}
\end{equation}

\noindent with $\tilde{\gamma}\rightarrow0$ and 
$\tilde{\epsilon}_l\rightarrow0$.

\section{Mean-field approximation}\label{MF}

In this section we specify the form of the variational function $Q(\phi)$. 
To allow for an analytical solution we neglect correlations between the 
group assignments and the remaining model parameters. We denote by $p_{ik}$ 
the probability that sample $i$ belongs to sample group $k$ and by 
$q_{jl}$ the probability that probe $j$ belongs to probe group $l$. 
Furthermore, given that $\theta$, $\pi$ and $\kappa$ always appear in 
different factors in (\ref{GM1}) or (\ref{GM2}) then their join 
distribution factorizes. Within the mean-field approximation for the group 
assignments and the later factorization the variational function can be 
written as

\begin{equation}\label{MF1}
Q(\phi) = \prod_i p_{ig_i} R(\theta)R(\pi)
\end{equation}

when dealing with the generative model (\ref{GM1}) and

\begin{equation}\label{MF2}
Q(\phi) = \prod_i p_{ig_i} \prod_j q_{jc_j} R(\theta)R(\pi)R(\kappa)
\end{equation}

\noindent when dealing with the generative model (\ref{GM2}), where $R(x)$ 
denotes a generic probability density function of $x$.

Summarizing, in the case studies below, we are going to solve the 
generative models (\ref{GM1}) or (\ref{GM2}), making use of renormalized 
invariant priors (Table \ref{priors}) and the MF variational function 
(\ref{MF1}) or (\ref{MF2}), respectively. This approach is based on the 
assumptions that: the population is divided in groups, the group 
assignments are generated by a multinomial model, the priors are 
renormalized invariant distributions, and a MF approximation of the 
variational solution with respect to the group assignments.

\section{Case study: Clustering real value data}\label{real}

Quite often we deal with datasets consisting of a real value measurement 
$X_{ij}$ over $i=1,\ldots,n$ samples and $j=1,\ldots,m$ variables, where 
the samples and variables are not necessarily independent. For simplicity, 
the particular kind of dependency we focus on is the existence of sample 
and variable groups. Our problem is to infer the sample and variable 
groups and the statistical parameters characterizing them.

To address this problem we consider the generative model (\ref{GM2}) with 
a normal likelihood, representing a two-sides Gaussian mixture model. The 
two-sides Gaussian mixture model is a natural extension of the Gaussian 
mixture model \cite{maclachlan00,rasmussen00} to characterize datasets 
with a group structure for both the samples and variables. Our 
contributions in this context are the use of prior distributions derived 
from symmetry arguments alone and the inclusion of a group structure at 
the variables side. The dataset, likelihood and priors associated with our 
statistical model are defined as follows:

{\em Data:} Consider $i=1,\ldots,n$ samples, $j=1,\ldots,m$ variables, and 
the real value measurements $X_{ij}$.

{\em Likelihood:} We assume that $X_{ij}$ are random variables with a 
normal distribution, with group dependent mean $\mu_{g_ic_j}$ and group 
independent variance $\sigma$, resulting in the likelihood

\begin{equation}\label{Preal}
P(X|g,c,\mu,\sigma) = \prod_{ij} \frac{1}{ \sqrt{2\pi\sigma^2} }
e^{ - \frac{\left(X_{ij}-\mu_{g_ic_j}\right)^2}{2\sigma^2} }\ .
\end{equation}

\noindent Here we are assuming that the main difference between groups is 
given by the means while the variance is group independent. The latter is 
a good approximation when the source of noise is given by the measurement 
itself and it behaves the same independently of the sample and variable 
group.

{\it Priors:} For the prior $P(\mu,\sigma)$ we generalize the Normal 
distribution prior in Table \ref{priors}. Accounting for more than one 
location parameter we obtain

\begin{eqnarray}\label{PGm}
P(\mu,\sigma) &=& \frac{ 2\left(\frac{\tilde{\alpha}}{2}\tilde{\sigma}^2\right)
^\frac{\tilde{\alpha}}{2} }{ \Gamma\left(\frac{\tilde{\alpha}}{2}\right)
\sigma^{\tilde{\alpha}+1} } e^{ -\tilde{\alpha}\frac{\tilde{\sigma}^2}{2\sigma^2} }\\
&\times& \prod_{kl} \sqrt{ \frac{\tilde{\alpha}}{2\pi\sigma^2} }
e^{ -\frac{\tilde{\alpha}}{2\sigma^2} \left(\mu_{kl}-\tilde{\mu}_{kl}\right)^2 }
\end{eqnarray}

\noindent and we work in the limit $\tilde{\alpha}\rightarrow0$.

To apply the variational method we consider the MF approximation 
(\ref{MF2}). Substituting the likelihood (\ref{Preal}), the priors 
(\ref{Ppi}), (\ref{Pkappa}) and (\ref{PGm}) and the MF variational 
function (\ref{MF2}) into (\ref{F}), and integrating over $\phi$ (summing 
over $g_i$ and $c_j$ and integrating over $\mu_{kl}$, $\sigma$, $\pi_k$ 
and $\kappa_l$) we obtain

\begin{eqnarray}\label{Freal}
F &\leq& {\rm const.} + (nm+KL+\tilde{\alpha}+1)\langle\ln\sigma\rangle
\nonumber\\
&+& \frac{1}{2} 
\sum_{ijkl}
p_{ik}q_{jl}\left( \langle\frac{1}{\sigma^2}\rangle X_{ij}^2 -
2X_{ij}\langle\frac{\mu_{kl}}{\sigma^2}\rangle
+ \langle\frac{\mu_{kl}^2}{\sigma^2}\rangle \right)
\nonumber\\
&+& \frac{\tilde{\alpha}}{2} \left[
\langle\frac{1}{\sigma^2}\rangle \tilde{\sigma}^2 
+ \sum_{kl} \left( \langle\frac{1}{\sigma^2}\rangle \tilde{\mu}_{kl}^2 
- 2\tilde{\mu}_{kl}\langle\frac{\mu_{kl}}{\sigma^2}\rangle
+ \langle\frac{\mu_{kl}^2}{\sigma^2}\rangle \right) \right]
\nonumber\\
&-&\sum_k\left(\sum_ip_{ik}+\tilde{\gamma}_k-1\right)
\langle\ln\pi_k\rangle
\nonumber\\
&-&\sum_l\left(\sum_jq_{jl}+\tilde{\epsilon}_l-1\right)
\langle\ln\kappa_l\rangle
\nonumber\\
&+& \int d\mu d\sigma R(\mu,\sigma)\ln R(\mu,\sigma)
\nonumber\\
&+& \int d\pi R(\pi)\ln R(\pi)
+ \int d\kappa R(\kappa)\ln R(\kappa)\kappa
\nonumber\\
&+&\sum_{ik} p_{ik}\ln p_{ik} + \sum_{jl} q_{jl}\ln q_{jl}
\end{eqnarray}

\noindent Minimizing (\ref{Freal}) with respect to $p_{il}$, $q_{jl}$, 
$R(\mu,\sigma)$, $R(\pi)$ and $R(\kappa)$ we obtain (VB-1):

\begin{equation}\label{preal}
p_{ik} = \frac{ e^{\langle\ln\pi_k\rangle -\frac{1}{2\sigma_*^2}
\sum_{jl} q_{jl}\left(
\frac{\sigma_*^2}{\alpha_{kl}} + \left(X_{ij}-\langle \mu_{kl}\rangle\right)^2
\right) } }
{\sum_s e^{ \langle\ln\pi_s\rangle -\frac{1}{2\sigma_*^2}
\sum_{jl} q_{jl}\left(
\frac{\sigma_*^2}{\alpha_{sl}} + \left(X_{ij}-\langle \mu_{sl}\rangle\right)^2
\right) } }
\end{equation}

\begin{equation}\label{qreal}
q_{jl} = \frac{ e^{ \langle\ln\kappa_l\rangle -\frac{1}{2\sigma_*^2}
\sum_{ik} p_{ik}\left(
\frac{\sigma_*^2}{\alpha_{kl}} + \left(X_{ij}-\langle \mu_{kl}\rangle\right)^2
\right) } }
{ \sum_s e^{ \langle\ln\kappa_l\rangle -\frac{1}{2\sigma_*^2}
\sum_{ik} p_{ik}\left(
\frac{\sigma_*^2}{\alpha_{ks}} + \left(X_{ij}-\langle \mu_{ks}\rangle\right)^2
\right) } }
\end{equation}

\begin{eqnarray}\label{Rmusigma}
R(\mu,\sigma) &=& \frac{ 2\left(\frac{\alpha}{2}\sigma_*^2\right)^{\frac{\alpha}{2}} }
{ \Gamma\left(\frac{\alpha}{2}\right) \sigma^{\alpha+1} }
e^{ -\frac{\alpha\sigma_*^2}{\sigma^2} }
\nonumber\\
&\times& \prod_{kl} \sqrt{ \frac{ \alpha_{kl} }{ 2\pi\sigma^2 } }
e^{ -\frac{\alpha_{kl}}{2\sigma^2} \left(\mu_{kl}-\langle\mu_{kl}\rangle\right)^2 }
\end{eqnarray}

\begin{equation}\label{alphakl}
\alpha_{kl} = \tilde{\alpha} + \sum_{ij}p_{ik}q_{jl}
\end{equation}

\begin{equation}\label{alpha1}
\alpha = \tilde{\alpha}+nm
\end{equation}

\begin{equation}\label{mu}
\langle\mu_{kl}\rangle = 
\frac{ \tilde{\alpha}\tilde{\mu}_{kl} + \sum_{ij} p_{ik}q_{jl}X_{ij} }
{ \tilde{\alpha} + \sum_{ij} p_{ik}q_{jl} }
\end{equation}

\begin{eqnarray}\label{sigma}
\sigma_*^2 &=& \frac{1}{\tilde{\alpha}+nm} \left[
\tilde{\alpha} \left( \tilde{\sigma}^2
+\sum_{kl}\left(\tilde{\mu}_{kl}^2
-\langle\mu_{kl}\rangle^2\right) \right) \right.
\nonumber\\
&+& \left. \sum_{ijkl}p_{ik}q_{jl} \left( X_{ij}^2-\langle\mu_{kl}\rangle^2 \right) \right]
\end{eqnarray}

\begin{equation}\label{P_pi}
R(\pi)={\rm D}(\pi;\gamma)\ ,\ \ \ \ 
\gamma_k = \tilde{\gamma}_k+\sum_ip_{ik}
\end{equation}

\begin{equation}\label{P_kappa}
R(\kappa)={\rm D}(\kappa;\epsilon)\ ,\ \ \ \
\epsilon_l = \tilde{\epsilon}_l+\sum_jq_{jl}
\end{equation}

\begin{eqnarray}\label{F_real}
F^* &=& {\rm const.} + \sum_{ik} p_{ik}\ln p_{ik} + \sum_{jl} q_{jl}\ln q_{jl}
- \ln{\rm B}(\gamma)
\nonumber\\
&-& \ln{\rm B}(\epsilon) + \frac{1}{2}\sum_{kl}\ln\alpha_{kl}\ .
\end{eqnarray}

\noindent These are a set of self-consistent equations which can be solved 
recursively to determine the probabilistic group assignments and the 
$\mu$, $\sigma$, $\pi$ and $\kappa$ distributions. They are the same in 
spirit as those for the EM algorithm \cite{dempster77}. Following 
\cite{mackay03,beal03} we refer to them as {\em variational Bayes} (VB) 
algorithm.

The main difference between the EM and VB algorithms is that in the former 
case we would take the average of the log likelihood over the group 
assignments but not over the distributions of $\mu$, $\sigma$, $\pi$ and 
$\kappa$. By taking the average over $\mu$ and $\sigma$ we obtain the 
additional $1/\alpha_{kl}$ term within the parenthesis in equations 
(\ref{preal}) and (\ref{qreal}). According to (\ref{alphakl}) $\alpha_{k}$ 
is equal to $\tilde{\alpha}$ plus the product of the average number of 
samples in sample group $k$ ($\sum_ip_{ik}$) and the average number of 
variables in variable group $l$ ($\sum_jq_{jl}$). Therefore, 
the $1/\alpha_{k}$ term penalizes assignments to small size groups. And it 
balances the contribution of $(X_{ij}-\langle\mu_{kl}\rangle)^2$, which 
drives the estimates towards a better fit and consequently groups of 
minimal size.

\subsection{VB implementation, real value data}

The actual implementation of the VB-1 algorithm in the context of real 
value data proceeds as follows. Set sufficiently large values for $K$ and 
$L$, larger than our expectation for the actual values of $K$ and $L$. In 
the following test examples we use $K=L=20$. Set the parameters 
$\tilde{\alpha}$, $\tilde{\mu}_{kl}$, $\tilde{\sigma}$, $\tilde{\gamma}_k$ 
and $\tilde{\epsilon}_l$. We set 
$\tilde{\alpha}=\tilde{\gamma}_k=\tilde{\epsilon}_l=10^{-6}$, 
$\tilde{\mu}_{kl}=0$ and $\tilde{\sigma}=1$. The choice of 
$\tilde{\mu}_{kl}$ and $\tilde{\sigma}$ is practically irrelevant provided 
we have chosen a sufficiently small $\tilde{\alpha}$. Set random initial 
conditions for $p_{ik}$ and $q_{jl}$. Starting from these random initial 
conditions iterate equations (\ref{preal})-(\ref{F_real}) until the 
solution converges up to some predefined accuracy. We use relative error 
of $F^*$ smaller than $10^{-6}$. In practice, compute 
$\langle\mu_{kl}\rangle$, $\alpha_{kl}$, $\sigma_*$, $\gamma_k$, 
$\langle\ln\pi_k\rangle$, $\epsilon_l$, $\langle\ln\kappa_l\rangle$, 
$p_{ik}$, $q_{jl}$ and $F^*$ in that order. To explore different potential 
local minima use different initial conditions and select the solution with 
lowest $F^*$. Since this algorithm penalizes groups with few members it 
turns out that, for sufficiently large $K$ and $L$, some sample and 
condition groups result empty. If this is not the case $K$ and/or $L$ 
should be increased until at least one sample group and one variable group 
results empty.

\begin{figure}[t]

\centerline{\includegraphics[width=3.2in]{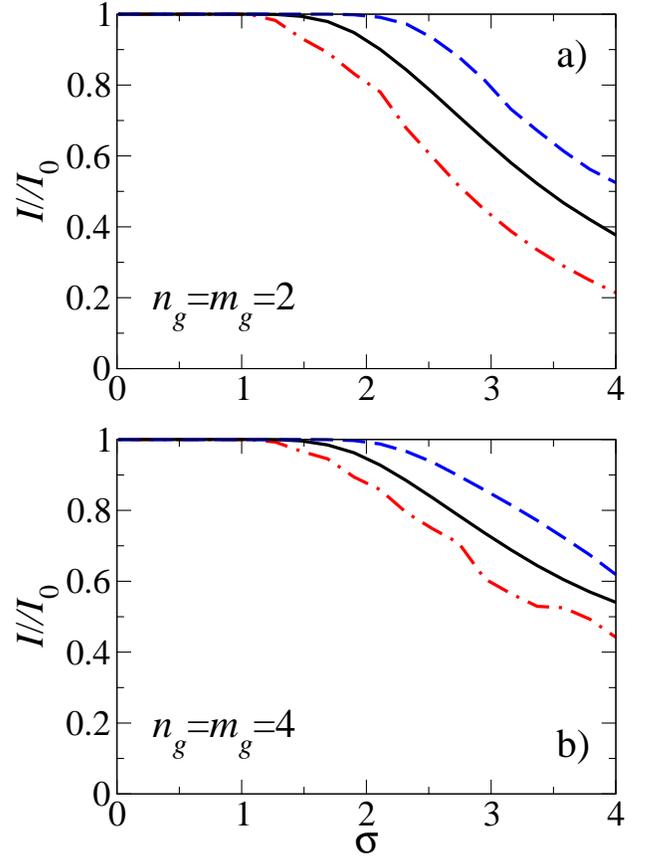}}

\caption{{\bf Clustering real value data:} Mutual information 
$I=I(p^O,p^*)$ between the original $p^O$ and estimated $p^*$ groups 
assignments, relative to its maximum value $I_0$ when $p^*=p^O$. The 
original data was made of $n=100$ samples divided in $K$ groups and 
$m=100$ conditions divided in $L$ groups. The values of $X_{ij}$ were 
extracted from a normal distribution with mean $\mu_{kl}=k+l$ and variance 
$\sigma$. The figure shows the mutual information between the original 
groups and the group assignment, estimated by the VB-1 algorithm, as a 
function of the variance $\sigma$. The dashed-dotted, solid and dashed 
lines corresponds with the worst, average and best case on 100 test 
examples, respectively. In a)  $K=L=2$ and in b) $K=L=4$. In both 
cases the mutual information is approximately equal to its maximum $I_0$ 
for values of $\sigma$ less than one, the minimum difference between the 
original means $\mu_{kl}$.}

\label{fig_real}
\end{figure}

\subsection{Test examples}

To test the performance of the VB-1 algorithm, 
(\ref{preal})-(\ref{F_real}), we consider test examples generated by 
the likelihood (\ref{Preal}) itself. Our aim is to test the variational 
result in the context of a relatively small number of samples and 
conditions. To quantify the goodness of the group assignment we consider 
the mutual information between the original $p^O$ 
($p^O_{ik}=\delta_{g_ik}$) and estimated $p^*$ sample group assignments,

\begin{equation}\label{Ip0p}
I(p^O,p^*) = \sum_{kk^\prime} \rho_{kk^\prime} \ln 
\frac{ \rho_{kk^\prime} }{ \rho^O_k\rho^*_{k^\prime} }
\end{equation}

\noindent where

\begin{equation}\label{rhopp}
\rho_{kk^\prime} = \frac{1}{n}\sum_i p^O_{ik} p^*_{ik}
\end{equation}

\begin{equation}\label{rhop0}
\rho^O_{k} = \frac{1}{n}\sum_i p^O_{ik}
\end{equation}

\begin{equation}\label{rhop}
\rho^*_k = \frac{1}{n}\sum_i p^*_{ik}\ .
\end{equation}

\noindent Note that $I(p^O,p^*)$ takes its maximum value when $p^*=p^O$, 
denoted by $I_0=I(p^O,p^O)$. Off course, the same could be done for the 
condition group assignments as well.

In our test examples the original data was made of $n=100$ samples divided 
in $K$ groups and $m=100$ conditions divided in $L$ groups. The values 
of $X_{ij}$ were extracted from a normal distribution with mean 
$\mu_{kl}=k+l$ and variance $\sigma$. We estimate the group assignment 
using the VB-1 algorithm, sampling one initial condition. Figure 
\ref{fig_real} shows the mutual information between the original and 
estimated groups as a function of the variance $\sigma$. In a) $K=L=2$ 
and in b) $K=L=4$. In both cases the mutual information is 
approximately equal to its maximum $I_0$ for values of $\sigma$ less than 
1. Since 1 is the minimum difference between the original means 
$\mu_{kl}$, we conclude that the VB-1 algorithm performs well when there 
is a significant difference between the distributions associated with 
different groups. For larger values of $\sigma$ the VB-1 algorithm 
performance starts to decrease. This is not, however, a deficiency of the 
algorithm but an unavoidable consequence of the mixing between the 
distributions coming from different groups. It is worth noticing that we 
obtain similar results for the case $K=4$ and $L=1$, indicating that 
the method works when there is no group structure on one side, in this 
case the conditions.

\section{Case study: clustering data represented by hypergraphs and 
bipartite graphs}

\begin{figure}[t]

\centerline{\includegraphics[width=3.2in]{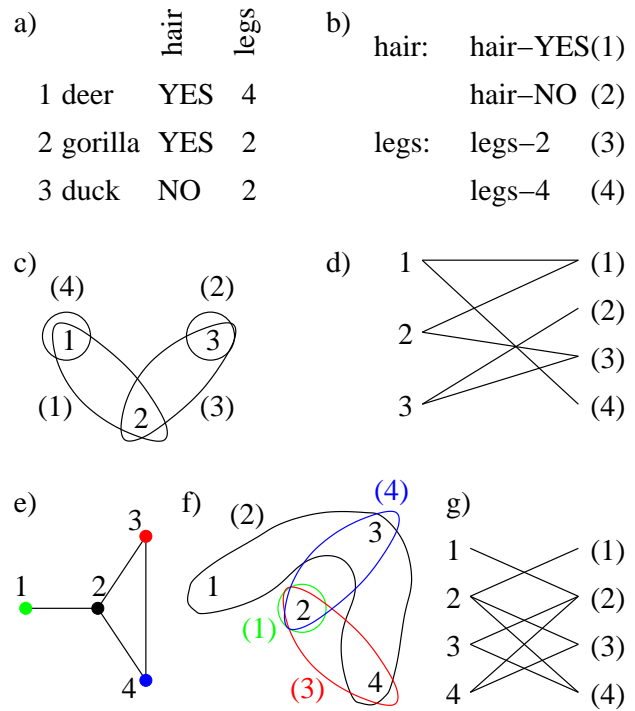}}

\caption{{\bf Hypergraph and bipartite graph data representations:} a) An 
example of a problem with categorical data. b) Mapping of the categorical 
variables onto augmented Boolean variables. c) Hypergraph representation 
of the categorical dataset in a). d) Bipartite graph representation of the 
categorical dataset in a). e) A graph example. f) Nearest-neighbor mapping 
of the graph in e) onto a hypergraph, where each hyper-edge represents a 
set of nearest neighbors of a vertex in the original graph, indicated by 
(1), (2), (3) and (4). g) Nearest-neighbor mapping of the graph in e) onto 
a bipartite graph. The original graph vertices are represented by 1, 2, 3 
and 4. The augmented bipartite graph vertices, representing 
nearest-neighbor sets, are represented by (1), (2), (3) and (4).}

\label{fig_hg_bg}
\end{figure}

There are several datasets consisting of a certain number of properties and 
the information of whether or not each sample exhibits each of the 
properties. For example, the dataset in Fig. \ref{fig_hg_bg} describes a 
population of three animals characterized by two attributes, hair and 
legs. The attribute hair can take the value YES (has hair) or NO (does not 
have hair) while the attribute legs takes the values 2 or 4 (at least 
within this dataset). The mathematical treatment of this problem is 
significantly simplified if the variables are mapped onto Boolean 
variables. To each $S$ states variable we associate $S$ Boolean variables, 
each representing the occurrence or not of a specific letter of the 
alphabet. For example, the attribute hair is associated with hair-YES and 
hair-NO and the attribute legs with legs-2 and legs-4 (Fig. 
\ref{fig_hg_bg}b). The outcome of this mapping is represented by the 
Boolean matrix $a_{ij}$, taking the value 1 if the answer to the Boolean 
variable $j$ is YES on sample $i$ and 0 otherwise.

Depending on our aim, the Boolean matrix can be represented either by a 
hypergraph or a bipartite graph. When we aim to cluster the samples 
without attempting to cluster the Boolean variables, $a_{ij}$ is better 
interpreted as the adjacency matrix of a hypergraph. A hypergraph is an 
intuitive extension of the concept of graph to allow for connections 
between more than two elements. In our case, the hypergraph vertices 
represent samples and hyper-edges, one associated which each Boolean 
variable, represent the set of all samples with the answer YES to the 
corresponding Boolean variable (Fig. \ref{fig_hg_bg}c). On the other hand, 
when we aim to cluster both the samples and Boolean variables then a 
bipartite graph interpretation is more appropriate, with one class of 
vertices for the samples and another one for the Boolean variables, and an 
edge connecting sample $i$ and variable $j$ whenever $a_{ij}=1$ 
(\ref{fig_hg_bg}d). The differences between these two approaches will 
become clear below.

\subsection{One side clustering: Statistical model on hypergraphs}

In this case the samples are assumed to be divided in groups while the 
hypergraph edges are modeled as independent. Here we follow the 
statistical model introduced in \cite{vazquez08}:

{\em Data:} Consider a hypergraph with a vertex set representing $n$ 
samples and $m$ edges characterizing the relationships among them. The 
hypergraph is specified by its adjacency matrix $a$, where $a_{ij}=1$ if 
element $i$ belongs to edge $j$ and it is 0 otherwise.

{\em Likelihood:} The adjacency matrix elements are generated by a 
binomial model with sample group and variable dependent probabilities 
$\theta_{kj}$, $k=1,\ldots,K$ and $j=1,\dots,m$, resulting in

\begin{equation}\label{Phg}
P(a|g,\theta) = \prod_{ij} \theta_{g_ij}^{a_{ij}}
\left(1- \theta_{g_ij}\right)^{1-a_{ij}}\ ,
\end{equation}

{\em Priors:} As priors we use the renormalized invariant prior of the 
binomial model (Table \ref{priors}). Taking into account that we have a 
binomial model for each pair of sample group and edge, we obtain

\begin{equation}\label{P_hg}
P(\theta) = \prod_{kj}{\rm Beta}(\theta_{kj};\tilde{\alpha}_{kj},\tilde{\beta}_{kj})
\end{equation}

\noindent with $\tilde{\alpha}_{kj}\rightarrow0$ and 
$\tilde{\beta}\rightarrow0$.

Substitute the likelihood (\ref{Phg}), the priors (\ref{Ppi}) and 
(\ref{P_hg}), and the MF variational function (\ref{MF1}) into (\ref{F}), 
and integrating over $\phi$ (summing over $g_i$ and integrating over 
$\theta_{kl}$ and $\pi_k$) we obtain

\begin{eqnarray}\label{F_hg}
F &\leq& - \sum_{jk} \left(\sum_ip_{ik}a_{ij}+\tilde{\alpha}_{kj}-1\right) 
\langle\ln\theta_{kj}\rangle
\nonumber\\
&-& \sum_{jk}\left(\sum_ip_{ik}(1-a_{ij})+\tilde{\beta}_{kj}-1\right)
\langle\ln(1-\theta_{kj})\rangle
\nonumber\\
&+& \sum_{ik} p_{ik}\ln p_{ik}
+ \int d\theta R(\theta)\ln R(\theta)
\nonumber\\
&+& \int d\pi R(\pi)\ln R(\pi)
+{\rm const.}
\end{eqnarray}

\noindent Minimizing (\ref{F_hg}) with respect to $p_{il}$,
$R(\theta)$ and $R(\pi)$ we obtain (VB-2)

\begin{equation}\label{p_hg}
p_{ik} = \frac{ e^{ \langle\ln\pi_k\rangle +
\sum_j \left[
a_{ij}\langle\ln\theta_{kj}\rangle
+(1-a_{ij})\langle\ln(1-\theta_{kj})\rangle
\right] } }
{ \sum_s e^{  \langle\ln\pi_s\rangle +
\sum_j \left[
a_{ij}\langle\ln\theta_{sj}\rangle
+(1-a_{ij})\langle\ln(1-\theta_{sj})\rangle
\right] } }
\end{equation}

\begin{equation}\label{Qtheta_hg}
R(\theta) = \prod_{kj} 
{\rm B}(\theta_{kj};\alpha_{kj},\beta_{kj})\ ,
\end{equation}

\begin{equation}\label{alpha_hg}
\alpha_{kj} = \tilde{\alpha}_{kj}+\sum_{ij}p_{ik}a_{ij}
\end{equation}

\begin{equation}\label{beta_hg}
\beta_{kj} = \tilde{\beta}_{kj}+\sum_{ij}p_{ik}(1-a_{ij})\ .
\end{equation}

\begin{equation}\label{Qpi_hg}
R(\pi)={\rm D}(\pi;\gamma)\ ,\ \ \ \ 
\gamma_k = \tilde{\gamma}_k + \sum_ip_{ik}
\end{equation}

\begin{eqnarray}\label{F_hg_min}
F^*&=& {\rm const.} +
\sum_{ik} p_{ik}\ln p_{ik}
\nonumber\\
&-& \sum_{kj}\ln {\rm B}(\alpha_{kj},\beta_{kj})
-\ln{\rm B}(\gamma)
\end{eqnarray}

\noindent These equations represent the VB algorithm for the statistical 
model on hypergraphs. In this case we have not been able to disentangle 
the contributions weighting the fit to the data and the model bias, both 
being included in the averages $\langle\ln(\theta_{kj})\rangle$ and 
$\langle\ln(1-\theta_{kj})\rangle$.

\subsection{VB algorithm implementation, statistical model on hypergraphs}

The implementation of the VB algorithm for the statistical model on 
hypergraphs proceeds as follows. Set sufficiently large values for $K$, 
larger than our expectation for the actual values of $K$. We use $K=20$ in 
the following test examples. Set the parameters $\tilde{\alpha}_{kj}$, 
$\tilde{\beta}_{kj}$ and $\tilde{\gamma}_k$. We set the parameters 
$\tilde{\alpha}_{kj}=\tilde{\beta}_{kj}=\tilde{\gamma}_k=10^{-6}$. Set 
random initial conditions for $p_{ik}$. Starting from these initial 
conditions iterate equations (\ref{p_hg})-(\ref{F_hg_min}) until the 
solution converges up to some predefined accuracy. We use relative error 
of $F^*$ smaller than $10^{-6}$. In practice, compute $\alpha_{kj}$, 
$\beta_{kj}$, $\langle\ln\theta_{kj}\rangle$, 
$\langle\ln(1-\theta_{kj})\rangle$, $\gamma_k$, $\langle\ln\pi_k\rangle$, 
$p_{ik}$ and $F^*$ in that order.  To explore different potential local 
minima use different initial conditions and select the solution with 
lowest $F^*$. Since this algorithm penalizes groups with few members it 
turns out that, for sufficiently large $K$, some sample and condition 
groups result empty. If this is not the case then increase $K$ until at 
least one group is empty. A matlab code implementing this algorithm can be 
found at http://www.sns.ias.edu/~vazquez/hgc.html.

\begin{figure*}[t]

\centerline{\includegraphics[width=5.8in]{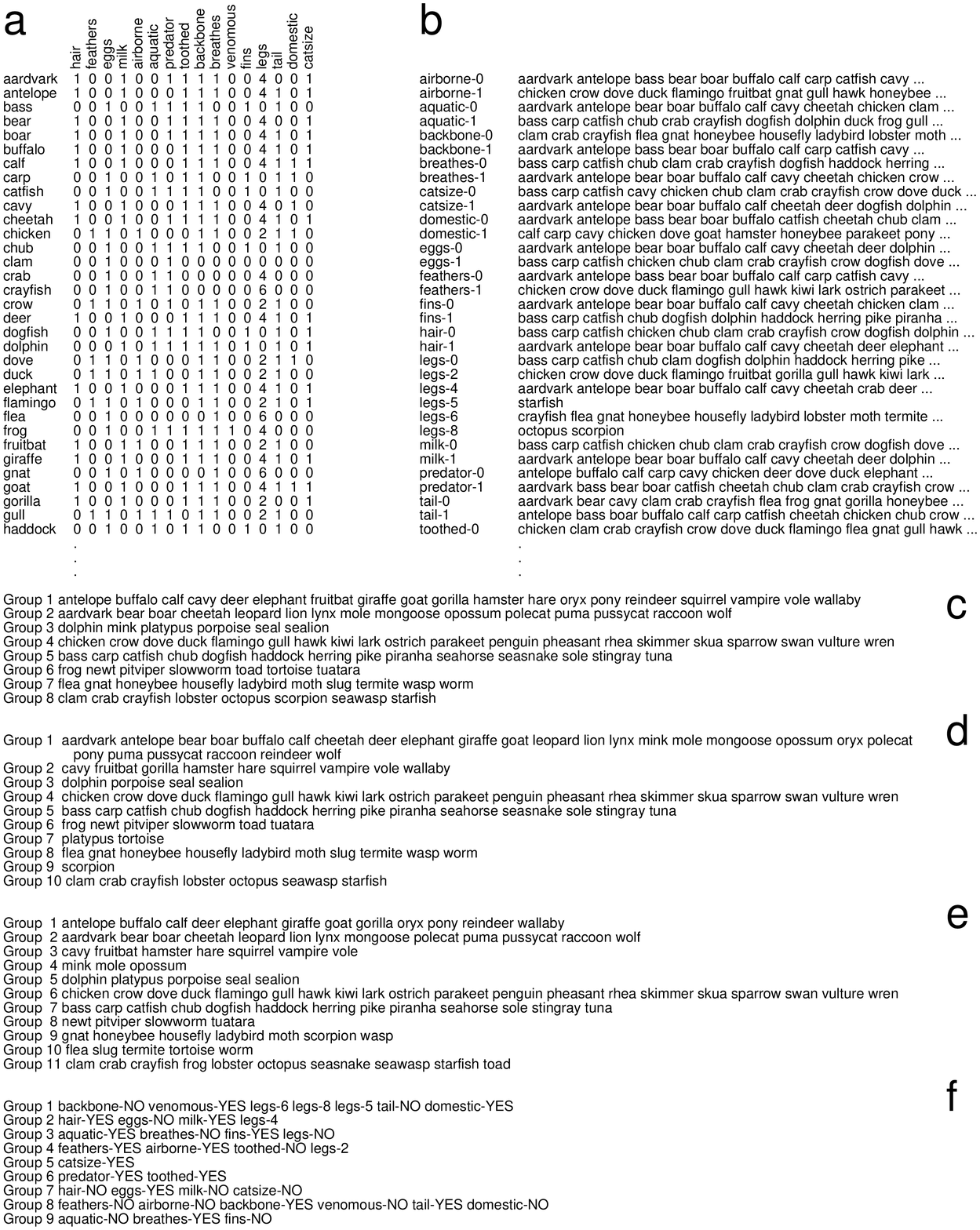}}

\caption{{\bf Stratification or an animal population:} a) A list of 
animals is given together with certain attributes characterizing them. The 
complete dataset is available from \cite{asuncion07}. Except for the 
attribute - legs - one and zero indicate possession or not, respectively, 
of the corresponding attribute. The problem consist on determining the 
optimal stratification of the animal population based on the provided 
attributes. b) Hypergraph representing the zoo data. Each line corresponds 
with an edge, whose elements are specified within the right column. c) 
Stratification as obtained in \cite{vazquez08}. d) Stratification by the 
VB-2 algorithm. e) and f) Stratification of the animal population e) and 
Boolean variables f) by the VB-3 algorithm.}

\label{fig_zoo}
\end{figure*}

\subsubsection{Test example: zoo problem}

Consider the animal population in Fig. \ref{fig_zoo}a together with their 
attributes: habitat, nutrition behavior, etc. Figure \ref{fig_zoo}b shows 
the mapping of this dataset onto a hypergraph. The hypergraph vertices 
represent animals and the edges represent the association between all 
animals with a given attribute: edge 1, all non-airborne animals; edge 2, 
all airborne animals, and so on.

The animal population stratification was already addressed in 
\cite{vazquez08}, finding the solution in Fig. \ref{fig_zoo}c. Although 
the starting statistical model is the same, the solution in 
\cite{vazquez08} was found assuming fixed the number of groups and 
estimating the group assignment using the EM algorithm (essentially a 
maximum likelihood estimate). Then, in an an attempt to focus in the 
solution with better consensus, solutions for different number of groups 
were obtained and the most representative solution was selected.

Here we address the same problem using a Bayesian approach and the 
variational solution. We start from the same statistical model on 
hypergraphs but now obtain a solution using the VB-2 algorithm 
(\ref{p_hg})-(\ref{F_hg_min}), sampling 10,000 initial conditions as in 
\cite{vazquez08}. The solution found by the VB-2 algorithm (Fig. 
\ref{fig_zoo}d) is quite similar to that previously found in 
\cite{vazquez08} (Fig. \ref{fig_zoo}c). The main differences are the 
splitting of the terrestrial mammals, the exclusion of the platypus and 
the tortoise from the amphibia-reptiles group and the scorpion from the 
terrestrial arthropods. More important, in both cases the main groups 
represent terrestrial mammals, aquatic mammals, birds, fishes, 
amphibia-reptiles, terrestrial arthropods and aquatic arthropods. The VB-2 
(\ref{p_hg})-(\ref{F_hg}) algorithm represents, however, a significant 
improvement over the approach followed in \cite{vazquez08}. It finds the 
consensus solution in one run, because it has built in the balance between 
better fitting and less bias.

\begin{figure}

\centerline{\includegraphics[width=3.2in]{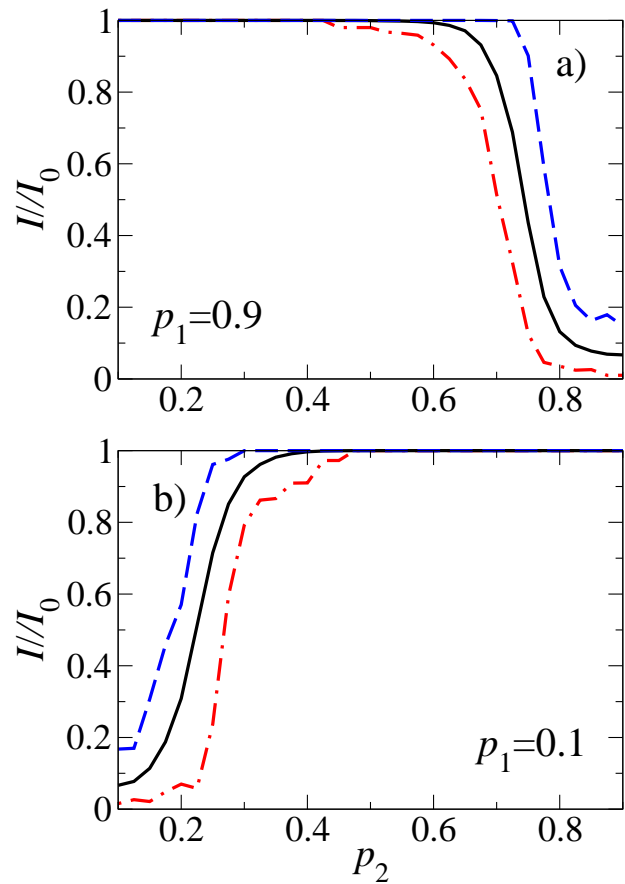}}

\caption{{\bf Finding graph modules, hypergraph model:} Mutual information 
$I=I(p^O,p^*)$ between the original $p^O$ and estimated $p^*$ groups 
assignments, relative to its maximum value $I_0$ when $p^*=p^O$. The 
original data was made of a graph with $n=100$ vertices divided in $K=2$ 
groups, with an intra- and inter-community connection probabilities $p_1$ 
and $p_2$, respectively. The figure shows the mutual information, between 
the original groups and the group assignment estimated by the VB-2 
algorithm (\ref{p_hg})-(\ref{F_hg_min}), as a function of the 
inter-community connectivity $p_2$. The dashed-dotted, solid and dashed 
lines corresponds with the worst, average and best case on 100 test 
examples. In a) we deal with dense communities ($p_1=0.9$)  and the 
algorithm performs well ($I/I_0\approx1$) for small values of the 
inter-community connectivity probability $p_2$.  In b) we deal with sparse 
communities ($p_1=0.1$) and the algorithm performs well for large values 
of the inter-community connectivity probability $p_2$.}

\label{fig_hg}
\end{figure}

\subsubsection{Test example: finding network modules}

The work by Newman and Leicht \cite{newman07} provides a hint on how to 
apply the hypergraph clustering to the problem of finding modules or 
communities in a graph or network. A graph is made by a set of vertices 
and a set of edges, the latter being pairs of connected vertices. The idea 
of Leicht and Newman is a ``guilty by association'' principle: vertices 
between the same module of a graph will tend to have connections to the 
same other vertices. This problem can be translated to a hypergraph 
problem, where the vertices are the graphs vertices, the hyper-edges are 
the set of nearest neighbors and the Boolean variables characterize 
whether or not a vertex belongs to the a set of nearest neighbors 
\cite{vazquez08} (Fig. \ref{fig_hg_bg}e and f). More precisely, to each 
vertex we associate a hyper-edge, given by the set of its nearest 
neighbors.  Therefore, there are $m=n$ hyper-edges, one for every vertex 
in the original graph. The hypergraph adjacency matrix has the matrix 
element $a_{ij}=1$ if vertex $i$ belongs to hyper-edge $j$, i.e. if vertex 
$i$ belongs to the nearest-neighbor set of vertex $j$, and $a_{ij}=0$ 
otherwise. If we label the nearest-neighbor sets with the same label as 
the vertices then the hypergraph adjacency matrix coincides with the 
adjacency matrix of the original graph. Thus, there is an exact mapping 
from the statistical model proposed by Newman and Leicht \cite{newman07} 
to the statistical model on hypergraphs.

Having specified this mapping we use the VB-2 algorithm 
(\ref{p_hg})-(\ref{F_hg_min}), sampling one initial condition, to find the 
graph modules in the original graph. To illustrate its performance we 
consider as a case study a graph composed by two communities, with 
probabilities $p_1$ and $p_2$ that two vertices within the same or 
different communities are connected, respectively. As already anticipated 
by Newman and Leicht \cite{newman07}, the nearest-neighbor approach can 
resolve both dense communities with lesser inter-community connections 
($p_1\gg p_2$) and sparse communities with more inter-community connections 
($p_1\ll p_2$).  Figure \ref{fig_hg} shows that the VB-2 algorithm 
performs quite well in those two regimes.

\subsection{Two sides clustering: statistical model on bipartite graphs}

We can face situations where there are groups of Boolean variables as well, requiring the 
clustering of both samples and Boolean variables. In this case the bipartite graph 
representation is more appropriate, with a class of vertices representing the samples and a 
class of vertices representing the Boolean variables. More precisely,

{\em Data:} Consider a bipartite graph with two vertex subsets, representing $n$ samples and 
$m$ Boolean variables. The graph is specified by its adjacency matrix $a$, where $a_{ij}=1$ 
when sample $i$ is connected to Boolean variable $j$, i.e. if Boolean variable $j$ is true for 
sample $i$, and $a_{ij}=0$ otherwise.

{\em Likelihood:} The adjacency matrix elements are generated by a binomial model with sample 
group and variable group dependent probabilities $\theta_{kl}$, $k=1,\ldots,K$ and 
$l=1,\dots,L$, resulting in

\begin{equation}\label{Pbg}
P(a|g,c,\theta) = \prod_{ij} \theta_{g_ic_j}^{a_{ij}}
\left(1- \theta_{g_ic_j}\right)^{1-a_{ij}}\ .
\end{equation}

{\em Priors:} For $P(\theta)$ we use the renormalized invariant prior of the 
binomial model. Taking into account that we have one binomial model per each pair 
of sample and variable group we obtain

\begin{equation}\label{P_bg}
P(\phi) =
\prod_{kj}{\rm B}(\theta_{kl};\tilde{\alpha}_{kl},\tilde{\beta}_{kl})
\end{equation}

\noindent with $\tilde{\alpha}_{kl}\rightarrow0$ and $\tilde{\beta}_{k}\rightarrow0$.

The likelihood (\ref{Pbg}) is quite similar to (\ref{Phg}), the main difference 
being that now the statistical properties of the Boolean variables appear through their 
corresponding group assignments $c_j$. This increases the model complexity by considering a 
group structure for the Boolean variables and, at the same time, reduces the number of 
$\theta$ parameters. Furthermore, (\ref{Pbg}) contains (\ref{Phg}) as the particular case 
where $L=n$ and one group associated to each Boolean variable.

Substituting the likelihood (\ref{Pbg}), the priors (\ref{P_bg}), (\ref{Ppi}) and 
(\ref{Pkappa}), and the MF variational function (\ref{MF2}) in (\ref{F}), and integrating over 
$\phi$ (summing over $g_i$ and $c_j$ and integrating over $\theta_{kl}$, $\pi_k$ and 
$\kappa_l$) we obtain

\begin{eqnarray}\label{F_bg}
F &\leq& - \sum_{kl} 
\left(\sum_{ij}p_{ik}q_{jl}a_{ij}+\tilde{\alpha}_{kl}-1\right) 
\langle\ln\theta_{kl}\rangle
\nonumber\\
&+& 
\sum_{kl}\left(\sum_{ij}p_{ik}q_{jl}(1-a_{ij})+\tilde{\beta}_{kl}-1\right)
\langle\ln(1-\theta_{kl})\rangle
\nonumber\\
&+& \sum_{ik} p_{ik}\ln p_{ik} +
\sum_{jl} q_{jl}\ln q_{jl}
\nonumber\\
&+& \int d\theta R(\theta)\ln R(\theta)
+ \int d\pi R(\pi)\ln R(\pi)
\nonumber\\
&+& \int d\kappa R(\kappa)\ln R(\kappa)
+{\rm const.}
\end{eqnarray}

\noindent Minimizing (\ref{F_bg}) with respect to $p_{il}$, $q_{jl}$,
$R(\theta)$, $R(\pi)$ and $R(\kappa)$ we obtain (VB-3)

\begin{equation}\label{p_bg}
p_{ik} = \frac{ e^{ \langle\pi_k\rangle +
\sum_{jl} q_{jl} \left[
a_{ij}\langle\ln\theta_{kl}\rangle
+(1-a_{ij})\langle\ln(1-\theta_{kl})\rangle
\right] } }
{ \sum_s e^{ \langle\pi_s\rangle +
\sum_{jl} q_{jl} \left[
a_{ij}\langle\ln\theta_{sl}\rangle
+(1-a_{ij})\langle\ln(1-\theta_{sl})\rangle
\right] } }
\end{equation}

\begin{equation}\label{q_bg}
q_{jl} = \frac{ e^{ \langle\kappa_l\rangle +
\sum_{ik} p_{ik} \left[
a_{ij}\langle\ln\theta_{kl}\rangle
+(1-a_{ij})\langle\ln(1-\theta_{kl})\rangle
\right] } }
{ \sum_s e^{ \langle\kappa_s\rangle +
\sum_{ik} p_{ik} \left[
a_{ij}\langle\ln\theta_{ks}\rangle
+(1-a_{ij})\langle\ln(1-\theta_{ks})\rangle
\right] } }
\end{equation}

\begin{equation}\label{Qtheta_bg}
R(\theta) = \prod_{kl} 
{\rm B}(\theta_{kl};\alpha_{kl},\beta_{kl})\ ,
\end{equation}

\begin{equation}\label{alpha_bg}
\alpha_{kl} = 1 + \sum_{ij}p_{ik}q_{jl}a_{ij}
\end{equation}

\begin{equation}\label{beta_bg}
\beta_{kl} = 1 + \sum_{ij}p_{ik}q_{jl}(1-a_{ij})\ .
\end{equation}

\begin{equation}\label{P_pi_bg}
R(\pi)={\rm D}(\pi;\gamma)\ ,\ \ \ \ 
\gamma_k = \tilde{\gamma}_k + \sum_ip_{ik}
\end{equation}

\begin{equation}\label{P_kappa_bg}
R(\pi)={\rm D}(\kappa;\epsilon)\ ,\ \ \ \ 
\epsilon_l = \tilde{\epsilon}_l + \sum_jq_{jl}
\end{equation}

\begin{eqnarray}\label{F_bg_min}
F^* &=& {\rm const.} +
\sum_{ik} p_{ik}\ln p_{ik} + \sum_{jl} q_{jl}\ln q_{jl}
\nonumber\\
&-& \sum_{kl}\ln {\rm B}(\alpha_{kl},\beta_{kl})
- \ln{\rm B}(\gamma) -\ln{\rm B}(\epsilon)
\end{eqnarray}

\noindent Equations (\ref{p_bg})-(\ref{F_bg_min}) represent the VB 
algorithm for the statistical model on bipartite graphs. They can be used 
to found modules or communities in graphs with a bipartite structure, 
including those representing samples and Boolean variables.

\subsection{VB algorithm implementation, statistical model on bipartite 
graphs}

The implementation of the VB-2 algorithm (\ref{p_bg})-(\ref{F_bg_min}) for the statistical 
model on bipartite graphs proceeds as follows. Set sufficiently large values for $K$ and $L$, 
larger than our expectation for the actual values of $K$ and $L$. Set the parameters 
$\tilde{\alpha}_{kl}$, $\tilde{\beta}_{kl}$, $\tilde{\gamma}_k$ and $\tilde{\epsilon}_l$. We 
set the parameters $\tilde{\alpha}_{kl} = \tilde{\beta}_{kl} = \tilde{\gamma}_k = 
\tilde{\epsilon}_l = 10^{-6}$. Set random initial conditions for $p_{ik}$ and $q_{jl}$. 
Starting from these initial conditions iterate equations (\ref{p_bg})-(\ref{F_bg_min}) until 
the solution converges up to some predefined accuracy. We use relative error of $F^*$ smaller 
than $10^{-6}$. In practice, compute $\alpha_{kj}$, $\beta_{kj}$, 
$\langle\ln\theta_{kj}\rangle$, $\langle\ln(1-\theta_{kj})\rangle$, $\gamma_k$, 
$\langle\ln\pi_k\rangle$, $\epsilon_l$, $\langle\ln\kappa_l\rangle$, $p_{ik}$, $q_{jl}$ and 
$F^*$ in that order. To explore different potential local minima use different initial 
conditions and select the solution with lowest $F^*$. Since this algorithm penalizes groups 
with few members it turns out that, for sufficiently large $K$ and $L$ some sample and/or 
variable groups result empty. If this is not the case, increase $K$ and/or $L$ until at least 
one sample group and one variable group results empty.

\subsubsection{Test example: zoo problem}

Let us go back to the zoo problem (Fig. \ref{fig_zoo}a). Now we represent 
this dataset by a bipartitite graph, with one class of vertices 
representing the animals and the other class the Boolean variables (e.g. 
Fig. \ref{fig_hg_bg}a,b and d) Using the VB-3 algorithm 
(\ref{p_bg})-(\ref{F_bg_min}), sampling 10,000 initial conditions as in 
\cite{vazquez08}, we perform a two-sides clustering of the bipartite graph 
obtaining the animal population stratification in Fig. \ref{fig_zoo}e and 
the Boolean variables stratification in Fig. \ref{fig_zoo}f. The animal 
clusters are similar to those previously obtained using the statistical 
model on hypergraphs (Fig. \ref{fig_zoo}c and d). The main difference is 
the more refined subdivision of terrestrial mammals, now split in four 
groups (1, 2, 3 and 4).

In addition to the animal population stratification the two-sides 
clustering provides association groups between the Boolean variables (Fig. 
\ref{fig_zoo}f). These associations reflect the fact that not all Boolean 
variables are independent, some of them are linked. For example, group 2 
cluster four typical attributes of terrestrial mammals, they have hair, do 
not put eggs, milk and have four legs. In the same way, group 3 clusters 
attributes of fishes and group four of birds. Thus, in general, the 
bipartite graph model and the resulting two-sides clustering provides more 
information than the hypergraph approach.

\begin{figure}

\centerline{\includegraphics[width=3.2in]{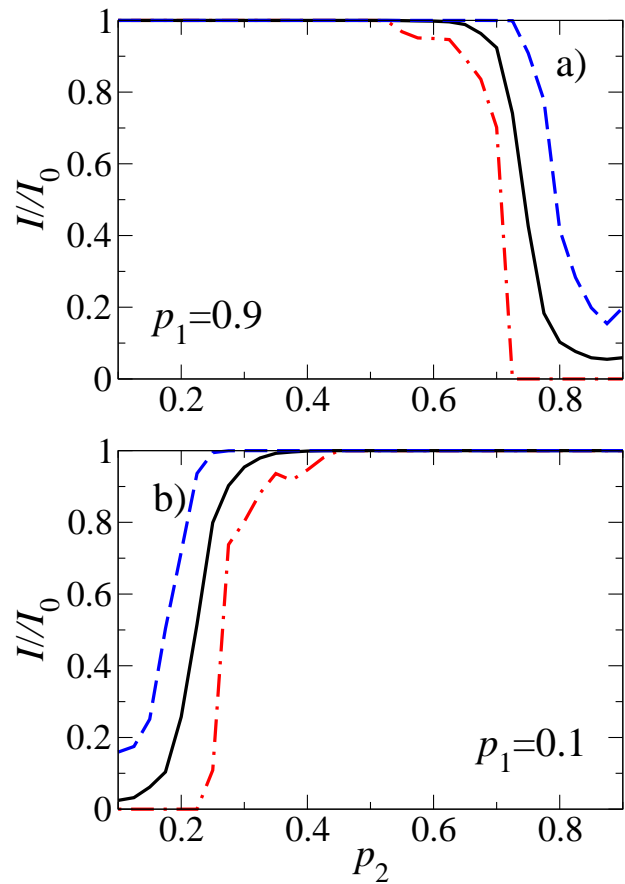}}

\caption{{\bf Finding graph modules, bipartite model:} Mutual information 
$I=I(p^O,p^*)$ between the original $p^O$ and estimated $p^*$ groups 
assignments, relative to its maximum value $I_0$ when $p^*=p^O$. The 
original data was made of a graph with $n=100$ vertices divided in $K=2$ 
groups, with an intra and inter-community connection probabilities $p_1$ 
and $p_2$, respectively. The figure shows the mutual information, between 
the original groups and the group assignment estimate by the VEM-3 
algorithm (\ref{p_bg})-(\ref{F_bg_min}), as a function of the 
inter-community connectivity $p_2$. The dashed-dotted, solid and dashed 
lines corresponds with the worst, average and best case on 100 test 
examples. In a) we deal with dense communities ($p_1=0.9$)  and the 
algorithm performs well ($I/I_0\approx1$) for small values of the 
inter-community connectivity probability $p_2$.  In b) we deal with sparse 
communities ($p_1=0.1$) and the algorithm performs well for large values 
of the inter-community connectivity probability $p_2$.}

\label{fig_bg}
\end{figure}

\subsubsection{Test example: finding network modules}

The bipartite graph model can be use to find network modules as well. In 
this case one class of vertices represents the original graph vertices and 
the other represents sets of nearest neighbors (Fig. \ref{fig_hg_bg}g). 
The two-sides clustering thus attempts to cluster both the original graph 
vertices and the sets of nearest neighbors. When the original graph is 
undirected the problem is symmetric (e.g. see Fig. \ref{fig_hg_bg}g). 
Indeed, if vertex $i$ belongs to the nearest-neighbor set of vertex $j$ 
then vertex $j$ belongs to the nearest-neighbor set of vertex $i$. As a 
consequence the clustering on the original vertices side cannot be 
differentiated from the clustering of nearest-neighbor sets. 
Intuitively this means that when two vertices belong to the same graph 
module we can say that their nearest-neighbor sets belong to the same 
nearest-neighbor set group.

Having specified this mapping we use the VB-3 algorithm 
(\ref{p_bg})-(\ref{F_bg_min}), sampling one initial condition, to find the 
graph modules in the original graph. To illustrate its performance we 
consider once again a graph composed by two communities, with 
probabilities $p_1$ and $p_2$ that two vertices within the same or 
different communities are connected, respectively. Figure \ref{fig_bg} 
shows that the VB-3 algorithm can resolve both dense communities with 
lesser inter-community connections ($p_1\gg p_2$) and sparse communities 
with more inter-community connections ($p_1\ll p_2$).

The comparison of Fig. \ref{fig_bg} and \ref{fig_hg} indicates that the 
bipartite graph model performs slightly better than the hypergraph model. 
For example, focusing on the average performance, for $p_1=0.9$ the VB-3 
algorithm performs almost perfectly till $p_2=0.6$, while the VB-2 
algorithm does till $p_2=0.5$. This could be, however, specific to the 
tested set of examples. Further research is required to determine which 
version performs better depending on the dataset under consideration.

\section{Discussion and conclusions}

The Bayesian approach allows for a systematic solution of data analysis 
problems. Its starting point is a statistical model of the data under 
consideration. From there, using Bayes rule, we can invert the statistical 
model to obtain the posterior distribution of the model parameters. The 
latter can be use, in principle, to calculate or compute averages or other 
magnitudes of interest.

One of the main criticisms to the Bayesian approach is the apparent 
ambiguity in selecting the prior distributions. Here we have worked 
further on Jaynes method \cite{jaynes68}, claiming that the prior 
distributions are given by the most general distribution dictated by the 
symmetries of the problem under consideration. One undesired consequence 
of this method is that when the symmetries are not sufficient constraints 
we obtain improper prior distributions. Yet, the use of improper priors 
can be avoided by working with renormalized distributions that are proper, 
and approach the improper prior in a certain limit. Using this approach we 
report here a correction to Jaynes prior for a likelihood with translation 
and scale invariance and a generalization of Jaynes prior for the binomial 
model to the multinomial model.

Having resolve the issue about the prior distributions, we can proceed to 
the application of the Bayesian approach to resolve a population 
structured. Taking inspiration from mixture models \cite{maclachlan00}, in 
particular Dirichlet mixture models \cite{blei03}, we introduce general 
statistical models with a built in population structure at the sample, and 
sample and variable, level. The model with a structure at the sample level 
aims one-side clustering problems, where the variables are assumed to be 
independent measurements. The model with a structure at both sample and 
variable level aims two-side clustering problems, where there are classes 
of variables. These statistical models are then postulated as generative 
models of some dataset. Introducing a MF approximation as variational 
function, we then resolve the population structure by solving the inverse 
problem, i.e. determining the sample and/or variable groups and model 
parameters from the data.

To illustrate the applicability and systematicity of the variational 
method, here we study the problem of data clustering, in the context of 
real value and Boolean variables. The outcome is a variational Bayes (VB) 
algorithm, a self-consistent set of equations to determine the group 
assignments and the model parameters. The VB algorithm is based on 
recursive equations similar to those for the EM algorithm, but with some 
intrinsic penalization for model bias. In the case of real value data, and 
under the assumption of normal distributions, the contributions favoring 
fitting and penalizing model bias are clearly disentangled. The fitting is 
quantified, as it is expected for normally distributed variables, by the 
mean square deviation. The model bias is quantified by the inverse of the 
square root of the mean cluster sizes. The tendency to reduce the mean 
square deviation is thus balanced by a tendency to increase the cluster 
sizes.

In the case of Boolean variables our analysis is based on a mapping into a 
hypergraph or bipartite graph. When we cluster the samples but not the 
Boolean variables the problem is mapped onto a statistical model on 
hypergraphs \cite{vazquez08}. On the other hand, when we perform a 
two-side clustering, clustering both the samples and the Boolean 
variables, the problem is mapped onto a statistical model on bipartite 
graphs.

The VB algorithms associated with the statistical model on hypergraphs and 
bipartite graphs can be used to find modules on a graph. Starting on an 
idea by Newman and Leicht \cite{newman07}, we show that the problem of 
graph modules can be mapped onto the problem of finding hypergraph modules 
or bipartite graph modules, where the hypergraph edges and the augmented 
bipartite graph vertices represent nearest-neighbor sets in the original 
graph. The resulting VB algorithms represent a significant improvement 
over the maximum likelihood approaches followed in \cite{newman07} and 
\cite{vazquez08}, by including a self-consistent correction for model 
complexity and bias.

It is worth mentioning that, depending on the starting statistical model, 
we could arrive to different versions of the VB algorithm. Indeed, for the 
finding graph modules problem we could use both the hypergraph and 
bipartite graph models. Furthermore, Hofman and Wiggins \cite{hofman07} 
have obtained another version based on a statistical model with different 
intra and inter-community connection probabilities. These approaches 
differ in the definition of what constitutes a group, community or module. 
We use the definition by Newman and Leicht \cite{newman07} based on 
topological similarity, i.e. two vertixes are topologically identical if 
they are connected to the same other vertices in the graph. Thus, we 
obtain group of vertices whose patterns of connectivity are similar. On 
the other hand, the definition used by Hofman and Wiggins \cite{hofman07} 
is based on the existence of two edge densities, characterizing the 
tendency of having an edge between intra- and inter-group pairs of 
vertices. Depending on the problem and the question we are asking we may 
adopt one or the other definition, and use the corresponding clustering 
method.

%\bibliography{network}

\end{document}